\documentclass[aps,prb,twocolumn,groupedaddress ]{revtex4-2}

\usepackage[T1]{fontenc}
\usepackage{amsmath}
\usepackage{amssymb}
\usepackage{amsfonts}
\usepackage{bbm}
\usepackage{xspace}
\usepackage{color}
\usepackage{graphicx}
\usepackage{ulem} 
\usepackage{placeins}
\usepackage{bm}
\usepackage{siunitx}
\usepackage{nccmath}
\usepackage{mathtools} 
\usepackage{empheq}
\usepackage{enumerate}
\usepackage{graphicx}
\usepackage{dcolumn}
\usepackage[caption=false]{subfig}

\usepackage[colorlinks=true,linkcolor=blue,citecolor=blue,urlcolor=blue]{hyperref}

\usepackage{MnSymbol}

\usepackage[version=4]{mhchem}

\graphicspath{{.}{Bilder/}}

\begin{document}

\title{Pseudospin-filter tunneling of massless Dirac fermions}
\author{Z.~D.~Li}
\author{W.~Zeng}
\email[E-mail: ]{zeng@ujs.edu.cn}
\affiliation{Department of physics, Jiangsu University, Zhenjiang 212013, China}


\begin{abstract}{}
The tunneling of the massless Dirac fermions through a vector potential barrier are theoretically investigated, where the vector potential can be introduced by the very high and very thin ($\delta$-function) magnetic potential barriers. We show that, distinct from the previously studied electric barrier tunneling, the vector potential barriers are more transparent for pseudospin-$1/2$ Dirac fermions but more obstructive for pseudospin-$1$ Dirac fermions. By tuning the height of the vector potential barrier, the pseudospin-$1/2$ Dirac fermions remain transmitted, whereas the transmission of the pseudospin-$1$ Dirac fermions is forbidden, leading to a pseudospin filtering effect for massless Dirac fermions.
\end{abstract}

\maketitle
\section{Introduction}\label{intro}
The pseudospin-$1/2$ massless Dirac fermions have been widely investigated since the graphene-like materials~\cite{wu2011valley,jiao2024transport,xu2023valley} were successfully fabricated in experiments. Novel phenomena have been found in pseudospin-$1/2$ massless Dirac system, such as the quantum Spin Hall effect~\cite{kane2005quantum,beeler2013spin,wang2020magnon}, specular Andreev reflection~\cite{beenakker2006specular}, Fabry-Pérot interference~\cite{li2022fabry,varlet2014fabry,mukherjee2018fabry}, Klein tunneling~\cite{wang2024valley} and laser irradiation induced pseudospin filter~\cite{martinez2009pseudospin}. Furthermore, the pseudospin-$1$ Dirac system featured by the flat band cutting through the Dirac point can also be realized in condensed matter system, such as the photonic crystals~\cite{fang2016klein,shen2010single}, Lieb~\cite{xu2014omnidirectional} and $\alpha-\mathcal{T}_3$~\cite{raoux2013dia} lattices. The dispersionless flat band leads to various unique phenomena, such as the super-Klein tunneling~\cite{singh2023geometrical,raoux2013dia,illes2017klein,cunha2021band,cunha2022tunneling,korol2021chiral,sukhachov2023stackings,betancur2017super}, super-Andreev reflection~\cite{zeng2022andreev,feng2020super} and Seebeck and Nernst effects~\cite{duan2023seebeck}.
\par
The electron tunneling through the electrostatic potential has been widely reported in the literature. It is shown that the electrons with different pseudospins exhibit different behaviors through the electrostatic potential tunneling process~\cite{yang2020effect,wang2010electronic,fang2019pseudospin,fan2013theoretical}. The Klein tunneling with the perfect transmission probability occurs only at the normal incidence for the pseudospin-$1/2$ Dirac fermions, whereas this perfect tunneling can be extended to omnidirection in pseudospin-$1$ Dirac systems, which is termed as the super-Klein tunneling~\cite{korol2021chiral,betancur2017super,sukhachov2023stackings}. Consequently, the electrostatic potential barriers are more transparent for the pseudospin-$1$ Dirac fermions than the pseudospin-$1/2$ Dirac fermions.
\par
However, the research on the electron tunneling through the vector potential barriers~\cite{lin2011gap,deng2013resonant,de2007magnetic,hu2019electron,yesilyurt2016klein,mandal2021tunneling} is insufficient. Motivated by this, here we theoretically study the electron tunneling through the vector potential barrier introduced by the very high and very thin ($\delta$-function) magnetic potential barriers. Both the pseudospin-$1/2$ and pseudospin-$1$ Dirac fermions are considered. We find that the transmission probabilities become zero for the pseudospin-$1$ Dirac fermions but nonzero for the pseudospin-$1/2$ Dirac fermions with the increasing of the magnetic potential, resulting in a pseudospin filtering effect for massless Dirac fermions.
\par
The paper is organized as follows. The model is presented in Section 2. The results and discussion are shown in Section 3. Finally, we conclude in Section 4.

\section{Model}\label{sec:2}
\begin{figure}[t]
    \centering
    \includegraphics[width=0.92\linewidth]{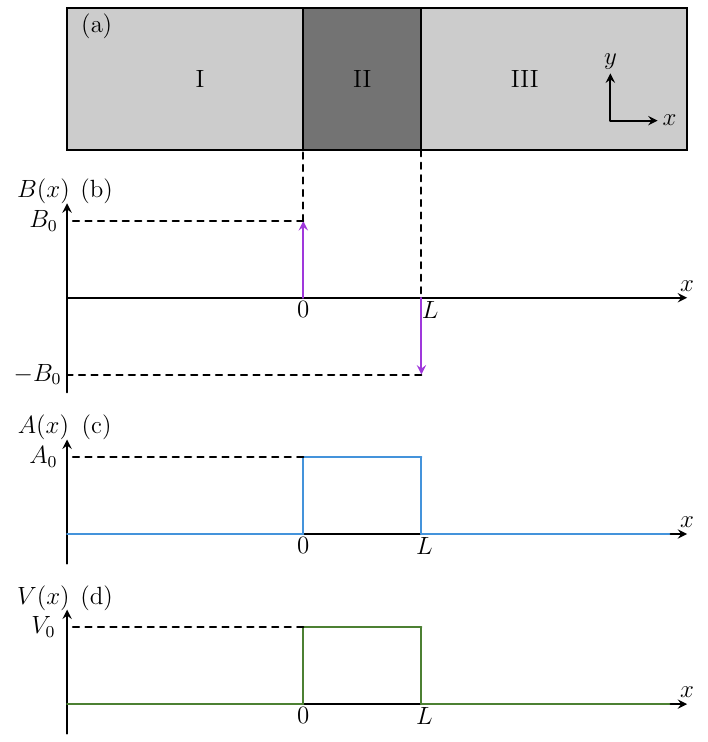}
    \caption{\footnotesize{(a) The schematic illustration of the two-dimensional tunneling junction in the $x-y$ plane where the transport direction is along the $x$-axis. (b) The profile of the $\delta$-function magnetic potential barriers located at $x=0$ and $x = L$. (c) The vector potential generated by the $\delta$-function magnetic barriers in Fig.\ \ref{mod}(b). (d) The electrostatic potential barrier with the magnitude of $V_{0}$.}}\label{mod}
\end{figure}

We consider the tunneling junction in the $x-y$ plane with the potential barrier in the region $0<x<L$, as shown in Fig.\ \ref{mod}. The out-of-plane magnetic potential is given by $\boldsymbol{B} = (0,0,B_{z})$ with
\begin{align}
    B_{z} = \ell_B B_{0}(\delta(x)-\delta(x-L)),
\end{align}
where $\ell_B = \sqrt{c/e B_0}$ is the magnetic length and $L$ is the width of the barrier region. The $\delta$-type magnetic potential barriers are located at the left and right end of the strip, with their magnitudes being $B_0$ and $-B_0$, respectively. The vector potential induced by $\boldsymbol{B} = \nabla \times \boldsymbol{A}$ reads $\boldsymbol{A} = (0,A_{y},0)$, where
\begin{align}
    A_{y} = \ell_B B_{0}(\Theta(x)-\Theta(x-L)).
\end{align}
The tunneling process is described by the low-energy Hamiltonian 
\begin{align}
    \mathcal{H} = v_F (\boldsymbol{p}+\frac{e}{c}\boldsymbol{A})\cdot \boldsymbol{\Sigma} + V(x),
\end{align}
where $\boldsymbol{p} = \hbar \boldsymbol{k}$ is the momentum with $\boldsymbol{k}$ being the wave vector, $V(x)$ is zero for the pristine regions (region I and region III in Fig.\ \ref{mod}(a)) and $V(x) = V_0$ for the barrier region (region II in Fig.\ \ref{mod}(a)), which can be adjusted by doping or by a gate voltage, and $\boldsymbol{\Sigma}$ is the Pauli matrix. For pseudospin-$1/2$ fermions, $\boldsymbol{\Sigma} = (\sigma_x, \sigma_y)$~\cite{wang2024valley,zalipaev2015resonant,allain2011klein}, where
\begin{align}\label{H1}
    \sigma_x = \begin{pmatrix}
        0 &1\\
        1& 0
    \end{pmatrix},
    \sigma_y = \begin{pmatrix}
        0 & -\mathrm{i}\\
        \mathrm{i} & 0
    \end{pmatrix}.
\end{align}
For pseudospin-$1$ fermions, $\boldsymbol{\Sigma} = (S_x, S_y)$\cite{zeng2022andreev,illes2017klein}, where
\begin{align}\label{H2}
    S_x = \frac{1}{\sqrt{2}} \begin{pmatrix}
        0 & 1 & 0 \\
        1 & 0 & 1 \\
        0 & 1 & 0
    \end{pmatrix},
    S_y = \frac{1}{\sqrt{2}} \begin{pmatrix}
        0 & -\mathrm{i} & 0 \\
        \mathrm{i} & 0 & -\mathrm{i} \\
        0 & \mathrm{i} & 0
    \end{pmatrix}. 
\end{align}
\par
The scattering states can be obtained by the schr{\"o}dinger equation
\begin{align}
    \mathcal{H}\psi(x)e^{ik_yy}=E\psi(x)e^{ik_yy},
\end{align}
where $E$ and $k_y$ are the incident energy and the conserved transverse wave vector, respectively.
The energy dispersion is given by
\begin{align}
    E = \hbar v_F \sqrt{k_x^2 + (k_y + \frac{1}{\ell_B})^2}.
\end{align}
The scattering basis states for pseudospin-$1/2$ are given by\begin{align}
    \psi^{+}_p = 
    \begin{pmatrix}
    1 \\
    e^{\mathrm{i}\theta} \end{pmatrix} e^{\mathrm{i} k_p x},\quad
        \psi^{-}_p = 
    \begin{pmatrix}
    1\\
    -e^{-\mathrm{i}\theta} \end{pmatrix} e^{ -\mathrm{i} k_p x},\label{eq:wf1}
\end{align}
and the scattering basis states for the pseudospin-$1$ Dirac fermions are given by
\begin{align}
    \psi^{+}_p = 
    \begin{pmatrix}
    e^{-\mathrm{i}\theta} \\
    \sqrt{2} \\
    e^{\mathrm{i}\theta} \end{pmatrix}
    e^{\mathrm{i} k_p x},\quad
        \psi^{-}_p = 
    \begin{pmatrix}
    -e^{\mathrm{i}\theta} \\
    \sqrt{2}\\
    -e^{-\mathrm{i}\theta} \end{pmatrix} e^{ -\mathrm{i} k_p x},\label{eq:wf2}
\end{align}
where $\theta$ is the incident angle satisfing $\sin\theta=\hbar v_Fk_y/(E-V_0)$ and $k_p=(E-V_0)\cos\theta/\hbar v_F$ is the longitudinal wave vector with $p = $ \{I, II, III\} distinguishing different regions of the junction. Consequently, the scattering states can be written as
\begin{align}\label{wf0}
    \psi(x) = \left\{
    \begin{matrix}
        \psi_I^+ + r\psi_I^-, &x<0\\
        a\psi_{II}^+ + b\psi_{II}^-, &0<x<L\\
        t\psi_{III}^+, &x>L
    \end{matrix}
    \right..
\end{align}
\par
The conservation of the current at the boundary requires the continuity of $(\psi_1,\psi_2)^T$ for the pseudospin-$1/2$ Dirac fermions and $(\psi_1+\psi_3,\psi_2)^T$ for the pseudospin-$1$ Dirac fermions, where $\psi_i$ ($i = 1,2,3$) is the $i$-th component of the wave function in Eq.\ (\ref{wf0})~\cite{illes2017klein,zhu2024transport}. The transmission and reflection probabilities are given by $T = \left|t\right|^2$ and $R = \left|r\right|^2$, respectively, which satisfy $R + T = 1$ due to the current conservation. One finds that the transmission probability reads
\begin{align}
    T &=\frac{1}{\cos^2(k_pL) + Z^2~\sin^2(k_pL)},\\
    Z &= \left\{
    \begin{matrix}
    \frac{1-~\sin\theta~\sin\theta'}{\cos\theta~\cos\theta'},&\text{pseudospin-$1/2$}\\
    \frac{\cos^2\theta'+~\cos^2\theta}{2~\cos\theta~\cos\theta'},&\text{pseudospin-$1$}        
    \end{matrix}\right.,
\end{align}
where the incident angle $\theta$ and the refraction angle $\theta'$ satisfy $\sin\theta =\hbar v_F k_y/E$ and $\sin\theta' = \hbar v_F (k_y+1/\ell_B)/(E-V_0)$, respectively.

The conductance can be calculated within the Landauer-B\"uttiker formalism~\cite{datta1992exclusion,cornean2005rigorous,pastawski1991classical}
\begin{align}
    G= G_0\int_{-\frac{\pi}{2}}^{\frac{\pi}{2}}\mathrm{d}\theta \ T\cos\theta,
\end{align}
where $G_0 = (2e^2/h)N(E)$ is the normalized conductance and $N(E) = (E - V_0)W/\pi \hbar v_F$ with $W$ being the junction width.

\section{Results}\label{sec:3}
\begin{figure}[t]
    \centering
    \includegraphics[width=1\linewidth]{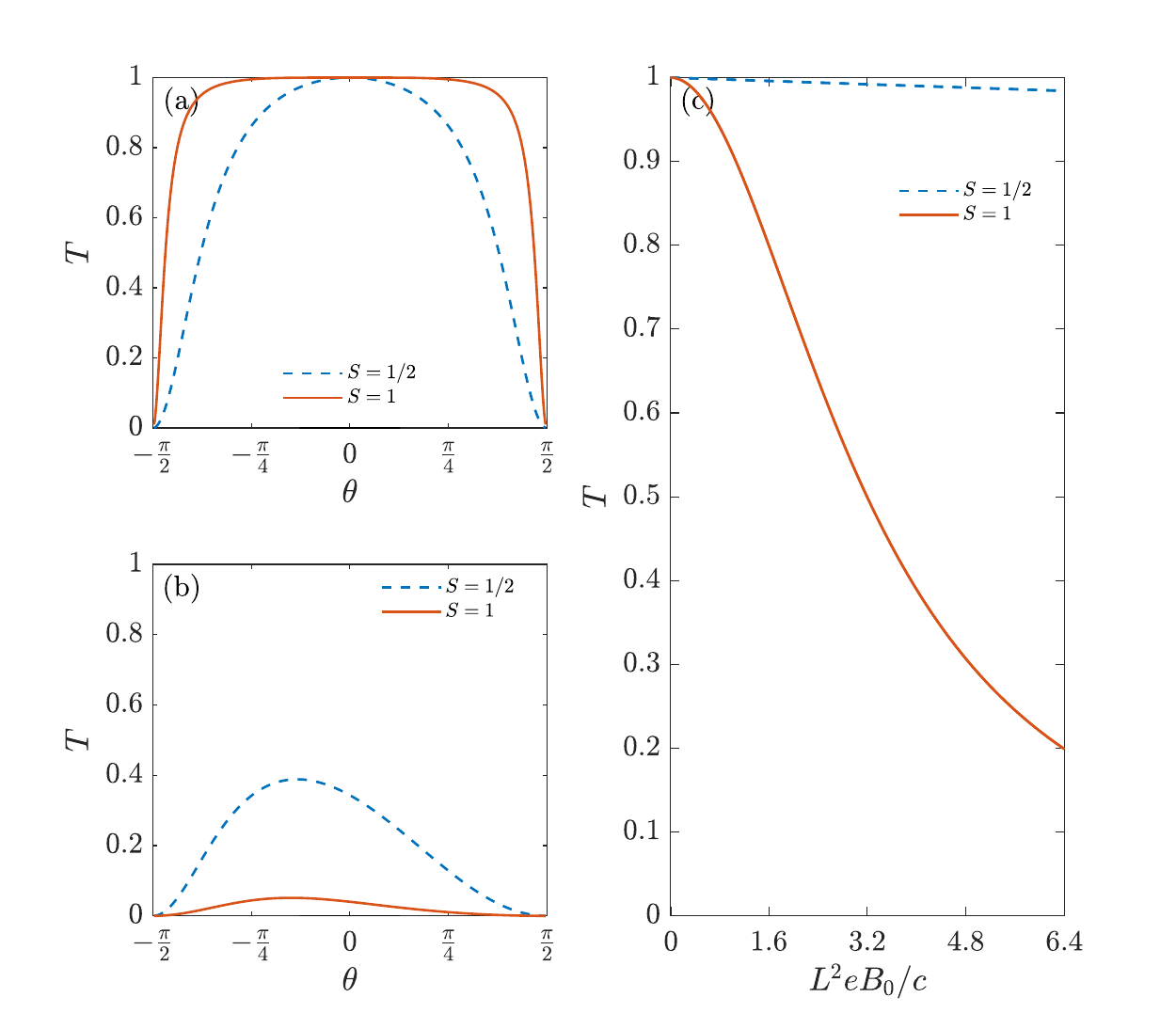}\\[5pt]
    \caption{\footnotesize{(a) The transmission probability as a function of the incident angle at $B_0 = 0$ with $E = 0.6V_0$. (b) The transmission probability as a function of the incident angle at $L^2 e B_0/c = 2$ with $E = 0.6V_0$. (c) The transmission probability at the normal incidence as a function of $B_0$ at $k_pL = 0.01$.}}\label{BTBN}
\end{figure}
\par 
In the absence of the magnetic field, the incident angle ($\theta$) resolved transmission probability ($T$) is shown in Fig.\ \ref{BTBN}(a), where the incident energy is $E = 0.6V_0$. It is shown that the perfect tunneling with the unit transmission probability for the pseudospin-$1/2$ Dirac fermions occurs at the normal incidence, \textit{i.e.}, $\theta = 0$, as the dashed blue line shown in Fig.\ \ref{BTBN}(a). However, for the pseudospin-$1$ Dirac fermions, the perfect transmission can be extended to wide incident angles around $\theta = 0$, namely, the super-Klein tunneling, as the solid red line shown in Fig.\ \ref{BTBN}(a). It is noted that the angle-resolved scattering pattern is symmetric due to the time-reversal symmetry, \textit{i.e.}, $T(\theta) = T(-\theta)$. 
\par
In the presence of the magnetic field, $T$ versus $\theta$ is shown in Fig.\ \ref{BTBN}(b). The transverse wave vector is shifted by $k_y\rightarrow k_y + 1/\ell_B$. As a result, the scattering pattern is asymmetric since the time-reversal symmetry is broken by the magnetic field, \textit{i.e.}, $T(\theta) \ne T(-\theta)$. The Klein tunneling is absent for both the pseudospin-$1/2$ and pseudospin-$1$ Dirac fermions. The maximum value of the transmission probability is about $0.4$ for the pseusospin-1/2 Dirac fermions. However, for the pseudospin-$1$ Dirac fermions, the maximum transmission probability is about $0.05$, which is much smaller than that for the pseudospin-$1/2$ Dirac fermions, indicating that, in contrast to the electrostatic potential barrier, the magnetic field induced vector potential barrier is more transparent for the pseudospin-$1/2$ Dirac fernions than the pseudospin-$1$ Dirac fermions.
\par
\begin{figure}[t]
    \centering
    \includegraphics[width=1\linewidth]{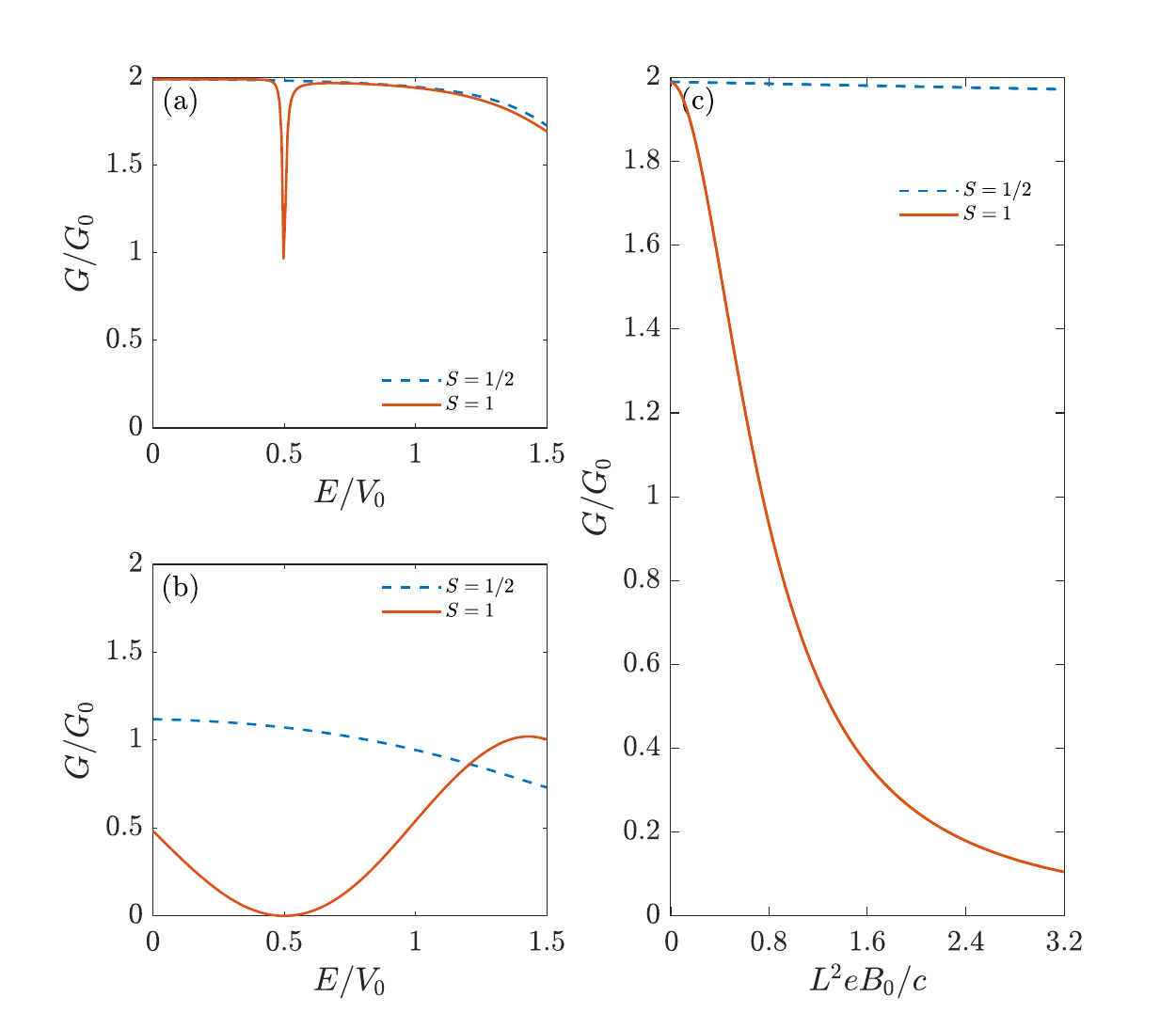}\\[5pt]
   \caption{\footnotesize{(a) The conductance as a function of the incident energy at $B_0 = 0$. (b) The conductance as a function of the incident energy at $L^2 e B_0/c = 2$. (c) The conductance as a function of $B_0$ at $k_pL = 0.01$.}}\label{BGBN}
\end{figure} 
\par
The transmission probability at the normal incidence as a function of the magnetic field $B_0$ is depicted in Fig.\ \ref{BTBN}(c) with $k_pL = 0.01$. It is shown that the transmission probability for the pseudospin-$1/2$ Dirac fermions ($T_{1/2}$) varies smoothly with the increasing of $B_0$. However, the transmission probability for the pseusospin-1 Dirac fermions ($T_1$) decreases rapidly with the increasing of $B_0$. Furthermore, the expressions of the transmission probability at the normal incidence can be obtained analytically as
\begin{gather}
    T_{1/2} = \frac{1}{\cos^2(k_pL)+\sec^2\theta_B\sin^2(2 k_p L)},\label{t12}\\
    T_{1} = \frac{4}{4\cos^2(k_p L)+(\cos\theta_B + \sec\theta_B)^2\sin^2(2 k_p L)},\label{t1}
\end{gather}
where $\sin\theta_B = 1/\ell_B(E-V_0)$ and $k_p = (E-V_0)\cos\theta_B/\hbar v_F$. For $k_pL \ll 1$, one finds Eq. (\ref{t12}) reads $T_{1/2} = 1/(1+(E-V_0)^2L^2/\hbar^2 v_F^2)$, which is independent of $B_0$. However, the $B_0$-dependent $T_1$ is given by $T_{1} = 4/(4+(\cos\theta_B+\sec\theta_B)(E-V_0)^2L^2/\hbar^2 v_F^2))$. For $B_0\rightarrow\infty$, we have $T_{1}/T_{1/2} \rightarrow 0$, as shown in Fig.\ \ref{BTBN}(c).
\par
The conductance as the function of the incident energy is shown in Figs.\ \ref{BGBN}(a) and \ref{BGBN}(b), where the magnetic field is zero in Fig.\ \ref{BGBN}(a) but nonzero in Fig.\ \ref{BGBN}(b). It is shown that in the presence of magnetic field, the conductance of the pseudospin-$1/2$ Dirac system is much greater than that of the pseudospin-$1$ Dirac system in the large range of the incident energy, which indicates that the vector potential barrier is more transparent for the pseudospin-$1/2$ Dirac fermions than the pseudospin-$1$ Dirac fermions. The conductance as a function of the magnetic field $B_0$ is shown in Fig.\ \ref{BGBN}(c), where $G$ becomes zero for the pseudospin-$1$ Dirac system but nonzero for the pseudospin-$1/2$ Dirac system for large $B_0$.

\section{Conclusions}\label{sec:4}
To conclude, we theoretically investigate the effect of the $\delta$-function magnetic potential barrier on the tunneling effect of Dirac fermions with different pseudospins under electrostatic potential. The pseudospin-$1$ Dirac fermions exhibits sensitivity to the magnetic potential, while the barrier appears more transparent for the pseudospin-$1/2$ Dirac fermions. By calculating and comparing the transmission probabilities and conductance of Dirac systems with different pseudospins, we verify the feasibility of utilizing magnetic potential to implement pseudospin filter.

\section*{Acknowledgements}
This work is supported by the College Student Innovation
Project (No. 202310299517X) and the Scientific Research Project of
Jiangsu University (No. 22A716).

\bibliography{ref}

\end{document}